# Machine-Checked Cardinality Bounds for Masked Barrett Reduction: A 1-Bit Side-Channel Leakage Barrier in Post-Quantum Cryptographic Hardware

**Ray Iskander**[1], **Khaled Kirah**[2,*]

[1] Verdict Security, ray@verdictsecurity.com
[2] Faculty of Engineering, Ain Shams University, Cairo, Egypt

## 1. Abstract

Barrett reduction is the nonlinear core of every practical NTT-based post-quantum cryptography implementation. Existing composition frameworks (ISW, t-SNI, PINI, DOM) address Boolean masking over GF(2); none provides a machine-checked characterization of Barrett's leakage under first-order arithmetic masking and the first-order probing model over prime fields. Building on our prior series, QANARY [1], partial-NTT-masking margins [2], algebraic foundations [3], and butterfly composition [4], we close this gap. We prove a trichotomy: for any $q > 0$ and shift $s$, the Barrett internal wire map $f_x(m) = ((x + 2^s - m) \bmod 2^s) \bmod q$ has preimage cardinality in $\{0,1,2\}$, never more. We call this the 1-Bit Barrier: max-multiplicity 2 implies at most 1 bit of min-entropy loss per internal wire, universal over all moduli. The count-zero cases, unreachable output values, reveal that actual leakage is often strictly less than 1 bit, making the bound conservative. We introduce PF-PINI (Prime-Field PINI): Barrett satisfies PF-PINI(2); the Cooley-Tukey butterfly satisfies PF-PINI(1). We observe (not yet proved) that with fresh inter-stage masking, the composed pipeline has max-multiplicity $\max(k_1, k_2)$, so the 1-Bit Barrier propagates. The trichotomy, the PF-PINI instantiations, and cardinality results are machine-checked in Lean 4 with Mathlib: 12 proved results, zero sorry, universal over all $q > 0$ (the min-entropy bound follows by standard definitions). Adams Bridge lacks fresh inter-stage masking, violating PF-PINI composition and explaining why Papers 1 [1] and 2 [2] found vulnerabilities. NIST IR 8547 recommends formal methods for PQC implementation validation. The 1-Bit Barrier provides the first universal machine-checked cardinality bound for masked Barrett reduction in ML-KEM (FIPS 203) and ML-DSA (FIPS 204), with a corresponding 1-bit leakage interpretation.

## 2. Introduction

Consider a hardware engineer designing a masked ML-KEM accelerator for FIPS 203 certification [5]. The NTT is implemented as a pipeline of Cooley-Tukey butterfly stages, each individually protected by first-order arithmetic masking with a fresh random mask. Between the butterfly stages,

*Correspondence Author: khaled.kirah@eng.asu.edu.eg
Ray Iskander: ray@verdictsecurity.com

Barrett reduction performs modular reduction, the step that keeps intermediate values in $\mathbb{Z}_q$. The engineer's question: is the full pipeline secure? For the butterfly stages, Paper 4 [4] answered yes, each output wire has exactly one mask producing each output value, a perfect bijection. For Barrett reduction, no prior machine-checked answer existed. Barrett is nonlinear: its conditional subtraction step breaks the affine structure that made the butterfly composition clean.

### 2.1. Why existing tools do not apply

The ISW probing model [6] and its refinements, t-SNI [7], PINI [8], DOM [9], provide powerful composition theorems for masked circuits. In their abstract formulation, these frameworks handle arbitrary rings. However, their machine-checked formalizations and verification tools (maskVerif [10], SILVER [11], Coco-Alma [12]) target Boolean circuits over GF(2). The NTT operates over $\mathbb{Z}_q$ with prime $q$. The composition frameworks that handle AND gates do not directly apply to modular reduction with carry chains and conditional correction steps. This is a scope gap in the existing formalization landscape, not a criticism of the frameworks themselves.

### 2.2. What this paper finds

The Barrett internal wire map, the function from mask value to wire observation, has a three-valued preimage structure: for any secret $x$ and any output value $v$, the number of masks producing $v$ is exactly 0, 1, or 2. Never more than 2. Universal over all $q > 0$ and all shift parameters $s$. The proof is three algebraic steps: (1) the preimage is a subset of two computable candidates $\{x - v, x - v + r\}$ where $r = (2^s : \mathbb{Z}_q)$; (2) each candidate contributes at most once; (3) a subset of a 2-element set has cardinality at most 2. The simplicity is the insight, not the proof itself, but what it reveals: Barrett's conditional subtraction creates exactly two preimage branches, both translations in $\mathbb{Z}_q$, and they overlap on at most one output value.

### 2.3. The support gap discovery

The count-zero cases deserve attention. When neither candidate $x - v$ nor $x - v + r$ is a valid preimage of $v$, the output value $v$ is unreachable for secret $x$ regardless of the mask. The support of the output distribution is smaller than $\mathbb{Z}_q$. This is not a security failure, it is a security feature. An unreachable output value cannot be used to distinguish secrets. The adversary's effective information is reduced, not increased, by the support gap. The 1-Bit Barrier is therefore a conservative upper bound: actual leakage is often strictly less than 1 bit. This framing matters because a reviewer encountering count-zero cases might conclude the masking is broken. It is not. Bounded max-multiplicity, not full-support uniformity, is the relevant security metric.

### 2.4. Contributions

This paper presents three results.

1. **The trichotomy theorem.** The Barrett internal wire map has preimage cardinality in $\{0,1,2\}$, universal over all $q > 0$ and all shift parameters $s$. Machine-checked in Lean 4 with zero `sorry`.

2. **The 1-Bit Barrier.** Max-multiplicity 2 implies $H_\infty(\text{output} \mid x) \geq \log_2(q) - 1$ per internal wire. For ML-KEM ($q = 3329$): $H_\infty \geq 10.70$ bits. For ML-DSA ($q = 8{,}380{,}417$): $H_\infty \geq 21.99$ bits.
3. **The PF-PINI framework.** Barrett satisfies PF-PINI(2). The identity masking gadget (modeling butterfly output wires) satisfies PF-PINI(1). We observe (not yet proved) that with fresh inter-stage masking, a composed pipeline has max-multiplicity $\max(k_1, k_2)$, the 1-Bit Barrier propagates.

### 2.5. Honest scope

The trichotomy proof is clean because of the two-branch algebraic structure of Barrett's conditional subtraction. The Lean theorems are universal over all $q > 0$, they do not even require $q$ to be prime. The scope condition $s \geq \lceil \log_2 q \rceil$, which ensures that $2^s \geq q$ and thus the Barrett approximation is valid, is satisfied by all practical implementations (Adams Bridge: $s = 24$, $q = 3329$, $2^{24} \gg 3329$). The composition observation, $\max(k_1, k_2)$ with fresh masking, is supported by the PF-PINI instantiations but is not a proved theorem. Higher-order masking ($d \geq 2$) and the PF-PINI composition proof are future work.

### 2.6. Adams Bridge connection

Adams Bridge [13] uses Barrett reduction with $s = 24$, $q = 3329$, satisfying the scope condition. Its 165 INSECURE_CONSERVATIVE Barrett wires from Paper 1 [1] correspond to internal wires that the trichotomy now characterizes: each leaks at most 1 bit per observation. However, Adams Bridge lacks fresh inter-stage masking (masking_en_ctrl = 1 only during rounds_count == 0; see ntt_ctrl.sv:264-272). The PF-PINI composition bound does not apply to Adams Bridge, which is precisely why Papers 1 [1] and 2 [2] found the vulnerabilities they did.

### 2.7. The five-paper program

Paper 1 [1] built the QANARY verification tool, scaling structural dependency analysis to 1.17 million cells across 30 Adams Bridge modules and identifying 165 INSECURE_CONSERVATIVE Barrett wires. Paper 2 [2] quantified the attack surface via belief propagation, showing 25–29 bits of security margin loss. Paper 3 [3] proved universal algebraic foundations in Lean 4: value-independence implies constant marginal distribution, with six theorems and zero sorry. Paper 4 [4] proved butterfly composition: each butterfly wire has exactly one mask producing each output (PF-PINI(1)), and $k$-stage pipelines with fresh masking preserve this property. Paper 5, this paper, closes the Barrett gap that Paper 4 explicitly deferred as Limitation (i), proving PF-PINI(2) for Barrett and establishing the 1-Bit Barrier as the universal leakage floor. The arc is complete: from tool construction through attack quantification, algebraic foundations, linear composition, to nonlinear characterization.

### 2.8. Organization

Section 2 reviews Barrett reduction and the masking model. Section 3 surveys related work. Section 4 proves the trichotomy theorem. Section 5 derives the 1-Bit Barrier security consequence. Section 6 defines the PF-PINI framework. Section 7 connects to Adams Bridge. Section 8 summarizes the proof suite. Section 9 discusses limitations. Section 10 concludes.

## 3. Background

### 3.1. Barrett Reduction in PQC Hardware

Barrett reduction computes $x \bmod q$ for an integer $x$ using precomputed constants, avoiding expensive division. The algorithm: multiply by an approximation $\mu = \lfloor 2^s/q \rfloor$, shift right by $s$, multiply by $q$, subtract, and apply one conditional correction. Table 1 lists the parameters used in Adams Bridge [13].

| Parameter | ML-KEM (FIPS 203) | ML-DSA (FIPS 204) |
|---|---|---|
| $q$ | 3329 | 8,380,417 |
| $\mu$ | 5039 | 33,587,228 |
| $s$ | 24 | 48 |
| $2^s \bmod q$ | 2385 | 196,580 |
| ROLLER | 767 | N/A |

**Table 1.** Barrett reduction parameters for Adams Bridge's ML-KEM (FIPS 203) and ML-DSA (FIPS 204) implementations [13]. The shift parameter $s$ satisfies the scope condition $2^s \geq q$ for both.

The conditional correction step, subtracting $q$ when the intermediate result exceeds $q$, is the source of nonlinearity. In Boolean masking, nonlinearity arises from AND gates. In Barrett reduction, it arises from a comparison-and-subtract operation that depends on the carry chain of the integer arithmetic.

Under first-order arithmetic masking, the fresh output mask rnd_24bit enters at the final output register. The output wire computes Barrett_internal$(x)$ − rnd_24bit, a translation that is trivially bijective (Paper 4's argument applies directly). The internal wires, those between the multiply, shift, and conditional subtraction steps, are the interesting case.

The **scope condition** is $2^s \geq q$. This is always satisfied in practice: standard Barrett reduction requires $s \geq 2\lceil \log_2 q \rceil$ for the approximation to be accurate. Adams Bridge uses $s = 24$ for $q = 3329$ ($2^{24} = 16{,}777{,}216 \gg 3329$). Notably, the Lean theorems we prove do not require this condition, they hold for all $s$, but the hardware relevance depends on it.

### 3.2. The Masking Model

First-order arithmetic masking over $\mathbb{Z}_q$ represents a secret $x$ as a pair of shares $(x - m, m)$ where $m$ is drawn uniformly from $\mathbb{Z}_q$. Under the standard probing model [6], an adversary observes one wire per clock cycle. For first-order security ($t = 1$), any single wire must have a distribution that does not depend on the secret.

For output wires, the structure is Barrett_internal$(x) - m$. The map $m \mapsto c - m$ is a bijection on $\mathbb{Z}_q$ for any constant $c$ (translation in a group). This is identical to Paper 4's butterfly wire analysis and provides perfect masking: each output value is produced by exactly one mask.

For internal wires, the hardware computes:

$f_x(m) = ((x + 2^s - m) \bmod 2^s) \bmod q$

This models unsigned $s$-bit subtraction: the $+2^s$ ensures the intermediate value is non-negative before the mod $2^s$ reduction. When $m \leq x$, the mod $2^s$ has no effect and the result is $(x - m) \bmod q$. When $m > x$, the subtraction wraps around $2^s$, and the result is $(x - m + r) \bmod q$ where $r = 2^s \bmod q$. This branch-dependent behavior is the source of the two-branch structure.

### 3.3. The QANARY Program Context

Paper 1 [1] built QANARY, an NTT-specific structural dependency analysis tool that scales to 1.17 million cells across 30 Adams Bridge modules. QANARY identified 198 Barrett wires in the ML-KEM Barrett module: 165 INSECURE_CONSERVATIVE (both shares interact without fresh masking) and 33 SECURE. Paper 2 [2] used belief propagation on NTT factor graphs to show that Adams Bridge's masking provides 25–29 fewer bits of security margin than a correctly masked implementation. Paper 3 [3] proved universal algebraic foundations in Lean 4: value-independence implies constant marginal distribution via an algebraic proxy MutualInfoZero, with six theorems and zero sorry, upgrading finite-domain Z3 [14] / cvc5 [15] checks to kernel-verified proofs.

Paper 4 [4] proved butterfly per-context uniformity, each butterfly wire has exactly one mask producing each output value and extended this to $k$-stage pipeline composition under fresh per-stage masking. Paper 4 explicitly stated Barrett reduction as Limitation (i): "The NTT butterfly is affine in the mask. Composition for nonlinear gadgets such as Barrett reduction remains future work." This paper closes that gap.

### 3.4. Lean 4 and Mathlib Infrastructure

All proofs are formalized in Lean 4 [16], version 4.30.0-rc1, using Mathlib [17] pinned at commit `322515540d7f`. Lean 4 is a dependently typed proof assistant whose kernel verifies every proof step. A theorem with zero `sorry` (Lean's placeholder for incomplete proofs) means every step has been kernel-checked.

We use `ZMod q` as the quotient ring $\mathbb{Z}/q\mathbb{Z}$, equipped with Mathlib's `CommRing`, `Fintype`, and `DecidableEq` instances. Preimage cardinalities are computed via `Finset.univ.filter` and `Finset.card`. The `linear_combination` and `omega` tactics handle the algebraic manipulations.

The artifact is self-contained: it depends only on Mathlib and does not import from the Paper 3 or Paper 4 artifacts. It references the definitions and results of Papers 3 [3] and 4 [4] conceptually, `WireFunction`, `ValueIndependentR`, `MutualInfoZero`, `ButterflyStage`, and `butterfly_wire_count_eq_one` are described for context, but the Barrett-specific definitions and proofs in this artifact are independent.

The term "zero `sorry`" means: the Lean kernel has verified every proof obligation. The trusted computing base is the Lean 4 kernel and the Lean compiler (for `native_decide` calls). This paper's artifact contains no `native_decide` calls, all proofs are kernel-verified algebraic arguments.

## 4. Related Work

### 4.1. Comparison with Prior Work

Table 2 summarizes how prior masking-verification approaches relate to this work along five axes: target masking model, Barrett-specific coverage, universality of the bound, machine-checked status, and PF-PINI applicability.

| Framework | Venue | Masking | Barrett | Universal | Machine-Checked | PF-PINI |
|---|---|---|---|---|---|---|
| ISW [6] | CRYPTO 2003 | Any ring (abstract) | No | Abstract | Pen-and-paper | No |
| t-SNI [7] | CCS 2016 | GF(2) | No | No | EasyCrypt | No |

| **Framework** | **Venue** | **Masking** | **Barrett** | **Universal** | **Machine-Checked** | **PF-PINI** |
|---|---|---|---|---|---|---|
| | | (tools) | | | | |
| maskVerif [10] | EUROCRYPT 2015 | GF(2) circuits | No | No | EasyCrypt | No |
| PINI [8] | TIFS 2020 | GF(2) (tools) | No | No | Pen-and-paper | No |
| DOM [9] | TIS@CCS 2016 | GF(2) circuits | No | No | No | No |
| SILVER [11] | ASIACRYPT 2020 | GF(2) circuits | No | No | BDD-based | No |
| REBECCA [18] | EUROCRYPT 2018 | GF(2) + glitches | No | No | No | No |
| Coco-Alma [12] | FMCAD 2021 | GF(2) circuits | No | No | No | No |
| Gigerl et al. [19] | ACNS 2023 | Arith. (A2B/B2A) | No | No | No | No |
| eVer [20] | ePrint 2026 | Arith. gadgets | No | No | No | No |
| Paper 3 [3] | arXiv 2026 | $\mathbb{Z}_q$ (Lean 4) | No | All $q > 0$ | Lean 4 | No |
| Paper 4 [4] | 2026 | $\mathbb{Z}_q$ (Lean 4) | No | All $q > 0$ | Lean 4 | PF-PINI(1) |
| **This work** | | $\mathbb{Z}_q$ **(Lean 4)** | **Yes** | **All** $q > 0$ | **Lean 4** | **PF-PINI(2)** |

***Table 2.*** *Prior masking-verification approaches and their coverage of Barrett reduction over* $\mathbb{Z}_q$*. Bold row marks the contribution of this paper.*

**Boolean composition frameworks.** The ISW transformation [6] provides $t$-probing security for arbitrary circuits. The t-NI and t-SNI notions [7] enable modular composition: t-SNI gadgets compose securely without global circuit analysis. Cassiers and Standaert's PINI [8] achieves trivially composable gadgets where each probe reveals at most one share index. The DOM methodology [9] provides pipeline-register-based composition for hardware implementations. These results are powerful, but their machine-checked formalizations and tool implementations (maskVerif [10], SILVER [11], Coco-Alma [12]) target Boolean circuits over GF(2). Bloem et al.'s REBECCA [18] extends to the glitch-robust probing model. In all cases, the formalization infrastructure targets Boolean masking.

**Arithmetic masking.** Gigerl et al. [19] extended formal masking verification to arithmetic operations, specifically A2B and B2A conversion gadgets at the individual gadget level. The eVer tool [20] addresses arithmetic gadget verification. Coron et al. [21] developed higher-order arithmetic masking conversion techniques. None of these works provides a machine-checked preimage bound for Barrett reduction over $\mathbb{Z}_q$, and none characterizes the preimage multiplicity structure of modular reduction under arithmetic masking.

**The specific gap.** ISW and t-SNI handle arbitrary rings in their abstract formulations, the gap is not that they are Boolean-only in theory. The gap is that no machine-checked formalization targets Barrett reduction specifically over $\mathbb{Z}_q$, none characterizes the preimage multiplicity structure (the trichotomy), and none provides a universal leakage bound that hardware designers can cite in FIPS 140-3 certification arguments without per-parameter-set re-verification.

## 5. The Trichotomy Theorem

This section contains the mathematical heart of the paper.

### 5.1. The Barrett Internal Wire Map

We define the Barrett internal wire map algebraically using the two-branch structure observed in hardware. For any $q > 0$, shift parameter $s$, secret $x \in \mathbb{Z}_q$, and mask $m \in \mathbb{Z}_q$:

$$f_x(m) = \begin{cases} x - m & \text{if } m.\text{val} \leq x.\text{val} \\ x - m + r & \text{otherwise} \end{cases}$$

where $r = (2^s : \mathbb{Z}_q)$ is the branch offset and $.val$ denotes the canonical natural number representative in $\{0, \ldots, q-1\}$.

In Lean 4:

```
/-- The Barrett internal wire map (algebraic two-branch form). -/
def barrettInternalMap {q : ℕ} [NeZero q] (s : ℕ)
   (x m : ZMod q) : ZMod q :=
  if m.val ≤ x.val then x - m else x - m + ↑(2 ^ s)
```

*Reproduced verbatim from artifact; compiles against Lean 4.30.0-rc1 + Mathlib 322515540d7f.*

The definition avoids integer underflow by working in $\mathbb{Z}_q$ directly: $x - m$ is computed in the ring, where subtraction is always defined. The branch condition $m.\text{val} \leq x.\text{val}$ determines whether the hardware's unsigned subtraction $x - m$ in $\{0, \ldots, 2^s - 1\}$ wraps around $2^s$. When it does not wrap (Branch A), the result is $x - m$. When it wraps (Branch B), the hardware adds $2^s$ to compensate, which in $\mathbb{Z}_q$ adds $r = 2^s \bmod q$.

The hardware-faithful Nat arithmetic definition is also provided for cross-validation:

```
/-- The Nat arithmetic definition (hardware-faithful). -/
def barrettInternalMapNat {q : ℕ} [NeZero q] (s : ℕ) (hs : q ≤ 2 ^ s)
   (x m : ZMod q) : ZMod q :=
  ↑((x.val + 2 ^ s - m.val) % 2 ^ s % q)
```

*Reproduced verbatim from artifact; compiles against Lean 4.30.0-rc1 + Mathlib 322515540d7f.*

The equivalence between barrettInternalMap and barrettInternalMapNat is computationally verified (all 3329 × 3329 input pairs for ML-KEM match) but not yet a proved Lean theorem. The algebraic definition is used for all proofs.

### 5.2. The Two-Branch Structure

The Barrett internal wire map always produces one of two values: a direct translation $x - m$ or a shifted translation $x - m + r$.

**Theorem 4.1** (barrettInternalMap_eq_or). *For any* $q > 0$, $s$, $x$, $m \in \mathbb{Z}_q$:

$$f_x(m) = x - m \quad \text{or} \quad f_x(m) = x - m + r$$

*where* $r = (2^s : \mathbb{Z}_q)$.

```
theorem barrettInternalMap_eq_or {q : ℕ} [NeZero q] (s : ℕ)
    (x m : ZMod q) :
    barrettInternalMap s x m = x - m ∨
    barrettInternalMap s x m = x - m + ↑(2 ^ s) := by
  unfold barrettInternalMap
  split
  · left; rfl
  · right; rfl
```

*Reproduced verbatim from artifact; compiles against Lean 4.30.0-rc1 + Mathlib 322515540d7f.*

The proof is immediate from the definition: the if expression produces exactly these two values. The key observation is that both branches are *translations* in $\mathbb{Z}_q$. In Branch A ($m$.val $\leq x$.val), the map is $m \mapsto x - m$. In Branch B ($m$.val $> x$.val), the map is $m \mapsto x - m + r$. Both are affine maps in $\mathbb{Z}_q$ (negation composed with translation), hence injective within their respective branches. This is the algebraic heart of the entire paper: Barrett's conditional subtraction creates exactly two injective branches, and their injectivity limits preimage sizes.

### 5.3. The Preimage Characterization

If a mask $m$ produces output $v$, then $m$ must be one of two computable candidates.

**Theorem 4.2** (barrettInternalMap_mem_pair). *If* $f_x(m) = v$, *then* $m = x - v$ *or* $m = x - v + r$.

```
theorem barrettInternalMap_mem_pair {q : ℕ} [NeZero q] (s : ℕ)
    (x m v : ZMod q) (hm : barrettInternalMap s x m = v) :
    m = x - v ∨ m = x - v + ↑(2 ^ s) := by
  rcases barrettInternalMap_eq_or s x m with h | h <;> [left; right] <;> {
    rw [h] at hm
    linear_combination -hm
  }
```

*Reproduced verbatim from artifact; compiles against Lean 4.30.0-rc1 + Mathlib 322515540d7f.*

The two candidates are: $x - v$ (the mask that produces $v$ via Branch A) and $x - v + r$ (the mask that produces $v$ via Branch B). These are the *only* two masks that can possibly produce output $v$ for secret $x$. The proof works by case-splitting on the two branches and solving the resulting linear equations via linear_combination.

**Theorem 4.3** (barrettInternalMap_preimage_subset). *The preimage set is contained in* $\{x - v, x - v + r\}$*:*

```
theorem barrettInternalMap_preimage_subset {q : ℕ} [NeZero q] (s : ℕ)
    (x v : ZMod q) :
    univ.filter (fun m : ZMod q => barrettInternalMap s x m = v) ⊆
    ({x - v, x - v + ↑(2 ^ s)} : Finset (ZMod q)) := by
  intro m hm
  simp only [mem_filter, mem_univ, true_and] at hm
  rw [mem_insert, mem_singleton]
  exact barrettInternalMap_mem_pair s x m v hm
```

*Reproduced verbatim from artifact; compiles against Lean 4.30.0-rc1 + Mathlib 322515540d7f.*

### 5.4. The 1-Bit Barrier

The main theorem follows directly: a subset of a 2-element set has cardinality at most 2.

**Theorem 4.4** (`barrett_max_multiplicity_two`). ***The 1-Bit Barrier.*** *For any* $q > 0$*, s,* $x$*,* $v \in \mathbb{Z}_q$*:*

$$|\{m \in \mathbb{Z}_q : f_x(m) = v\}| \leq 2$$

```
theorem barrett_max_multiplicity_two {q : ℕ} [NeZero q] (s : ℕ)
    (x v : ZMod q) :
    (univ.filter (fun m : ZMod q =>
      barrettInternalMap s x m = v)).card ≤ 2 := by
  calc (univ.filter (fun m => barrettInternalMap s x m = v)).card
      ≤ ({x - v, x - v + ↑(2 ^ s)} : Finset (ZMod q)).card :=
        card_le_card (barrettInternalMap_preimage_subset s x v)
    _ ≤ 2 := by
        have h := card_insert_le (x - v) ({x - v + ↑(2 ^ s)} : Finset (ZMod q))
        rw [card_singleton] at h; exact h
```

*Reproduced verbatim from artifact; compiles against Lean 4.30.0-rc1 + Mathlib 322515540d7f.*

The proof is a calc chain: the preimage is a subset of a 2-element set (Theorem 4.3), and a 2-element set has cardinality at most 2 (card_insert_le + card_singleton). This is the three-step argument promised in the introduction: preimage ⊆ two candidates, each candidate contributes at most once, subset of 2-element set has at most 2 elements.

**Theorem 4.5** (barrett_multiplicity_trichotomy). *The preimage cardinality is exactly 0, 1, or 2:*

```
theorem barrett_multiplicity_trichotomy {q : ℕ} [NeZero q] (s : ℕ)
    (x v : ZMod q) :
    let count := (univ.filter (fun m : ZMod q =>
      barrettInternalMap s x m = v)).card
    count = 0 ∨ count = 1 ∨ count = 2 := by
  have h := barrett_max_multiplicity_two s x v; omega
```

*Reproduced verbatim from artifact; compiles against Lean 4.30.0-rc1 + Mathlib 322515540d7f.*

The trichotomy follows immediately: the count is a natural number bounded above by 2 (Theorem 4.4) and below by 0 (trivially), so it must be 0, 1, or 2.

### 5.5. The Support Gap Discovery

When the preimage count is zero, the output value $v$ is unreachable for secret $x$: no mask produces it. This means the support of the output distribution, the set of values that can actually appear on the wire, is strictly smaller than $\mathbb{Z}_q$.

Table 3 shows representative computational evidence for ML-KEM ($q = 3329, s = 24, r = 2385$).

| Secret $x$ | Unreachable values (count = 0) | Doubled values (count = 2) |
|---|---|---|
| 0 | 1 | 1 |
| 100 | 101 | 101 |
| 832 ($\approx q/4$) | 833 | 833 |
| 1664 ($\approx q/2$) | 944 | 944 |
| 2496 ($\approx 3q/4$) | 832 | 832 |
| 3327 ($q - 2$) | 1 | 1 |
| 3328 ($q - 1$) | 0 (bijective) | 0 |

**Table 3.** Support-gap structure of the Barrett internal wire map for selected secrets in ML-KEM ($q = 3329, s = 24, r = 2385$). The unreachable count equals the doubled count by conservation. At $x = q - 1$ the map is a bijection.

The structure is symmetric: the number of unreachable values equals the number of doubled values (because the total count must sum to $q$). At $x = q - 1$, the map is a bijection, every output value is produced by exactly one mask.

**Observation 4.1** (Support gap formula). *The number of unreachable values for secret* $x$ *is* $\min(x.\text{val} + 1,\ q - r,\ q - 1 - x.\text{val})$ *where* $r = 2^s \bmod q$. The gap grows linearly from 0 (at $x = q - 1$) to a plateau of $q - r$ (in the middle range), then shrinks back to 0. This formula is computationally verified across 14 primes (including all PQC standard moduli) but is not a proved Lean theorem. The ZMod.val arithmetic required is technically tractable but not yet formalized.

**Security interpretation.** The support gap is a security *feature*, not a defect. An adversary observing an internal wire sees a distribution over at most $q$ values. When some values are unreachable, the adversary learns that those values will never appear, but this information is about the output distribution, not about the secret. The adversary's distinguishing advantage is bounded by the max-probability over *reachable* values, which is at most $2/q$ (Theorem 4.4). The unreachable values provide no additional attack surface. The 1-Bit Barrier is therefore a conservative bound: when the support gap is large, the actual min-entropy exceeds $\log_2(q) - 1$.

### 5.6. Tightness

The bound is tight: there exist $x$ and $v$ such that exactly two masks produce $v$.

**Theorem 4.6** (barrett_count_eq_two). *When both candidates hit* $v$ *and* $r \neq 0$ *in* $\mathbb{Z}_q$, *the preimage has exactly 2 elements:*

```
theorem barrett_count_eq_two {q : ℕ} [NeZero q] (s : ℕ)
  (x v : ZMod q)
  (hA : (x - v).val ≤ x.val)
  (hB : ¬ (x - v + ↑(2 ^ s)).val ≤ x.val)
```

```
    (hr : (↑(2 ^ s) : ZMod q) ≠ 0) :
    (univ.filter (fun m : ZMod q =>
      barrettInternalMap s x m = v)).card = 2 := by
  -- The two candidates are distinct (since r ≠ 0)
  have hne : x - v ≠ x - v + ↑(2 ^ s) :=
    fun heq => hr (by linear_combination -heq)
  -- Both candidates are in the preimage
  have hm1 : x - v ∈ univ.filter (fun m => barrettInternalMap s x m = v) :=
    mem_filter.mpr ⟨mem_univ _, barrettInternalMap_candidate_A s x v hA⟩
  have hm2 : x - v + ↑(2 ^ s) ∈ univ.filter (fun m => barrettInternalMap s x m =
v) :=
    mem_filter.mpr ⟨mem_univ _, barrettInternalMap_candidate_B s x v hB⟩
  -- The pair {m₁, m₂} ⊆ preimage, with card = 2
  have pair_sub : ({x - v, x - v + ↑(2 ^ s)} : Finset (ZMod q)) ⊆
      univ.filter (fun m => barrettInternalMap s x m = v) := by
    intro m hm
    rw [mem_insert, mem_singleton] at hm
    rcases hm with rfl | rfl <;> assumption
  have pair_card : ({x - v, x - v + ↑(2 ^ s)} : Finset (ZMod q)).card = 2 :=
    card_pair hne
  -- 2 ≤ count ≤ 2, so count = 2
  have h_ge : 2 ≤ (univ.filter (fun m => barrettInternalMap s x m = v)).card :=
by
    rw [← pair_card]; exact card_le_card pair_sub
  have h_le := barrett_max_multiplicity_two s x v
  omega
```

*Reproduced verbatim from artifact; compiles against Lean 4.30.0-rc1 + Mathlib 322515540d7f.*

The condition $r \neq 0$ holds for all odd primes $q$: since $\gcd(q, 2) = 1$, we have $q \nmid 2^s$, so $2^s \not\equiv 0 \pmod{q}$. For $q = 3329$, $s = 24$: the witness $x = 100$, $v = 0$ gives count = 2 (both candidates satisfy their branch conditions). The bound max-multiplicity $\leq 2$ is not just an upper bound with slack, it is achieved.

The supporting lemmas barrettInternalMap_candidate_A and barrettInternalMap_candidate_B verify that each candidate produces the correct output under its respective branch condition:

```
theorem barrettInternalMap_candidate_A {q : ℕ} [NeZero q] (s : ℕ)
    (x v : ZMod q) (hA : (x - v).val ≤ x.val) :
    barrettInternalMap s x (x - v) = v := by
  unfold barrettInternalMap; rw [if_pos hA]; ring

theorem barrettInternalMap_candidate_B {q : ℕ} [NeZero q] (s : ℕ)
    (x v : ZMod q) (hB : ¬ (x - v + ↑(2 ^ s)).val ≤ x.val) :
    barrettInternalMap s x (x - v + ↑(2 ^ s)) = v := by
  unfold barrettInternalMap; rw [if_neg hB]; ring
```

*Reproduced verbatim from artifact; compiles against Lean 4.30.0-rc1 + Mathlib 322515540d7f.*

## 6. The 1-Bit Barrier Security Consequence

### 6.1. Max Output Probability

The max-multiplicity bound directly implies a probability bound.

**Definition.** The max output probability for secret $x$ and output $v$ is:

$$\Pr[\text{output} = v \mid \text{secret} = x] = \frac{|\{m \in \mathbb{Z}_q : f_x(m) = v\}|}{q}$$

since each of the $q$ equally-likely masks contributes equally.

**Observation 5.1** (Max output probability bound). *For any $x$, $v$:*

$$\Pr[\text{output} = v \mid x] \leq \frac{2}{q}$$

*Proof.* The numerator is at most 2 (Theorem 4.4). The denominator is $q$. Therefore $\Pr[\text{output} = v \mid x] \leq 2/q$.

This bound is universal: it holds for any $q > 0$, any $s$, any $x$, and any $v$.

### 6.2. The Min-Entropy Bound

**Observation 5.2** (The 1-Bit Barrier, information-theoretic form). *For any $q > 0$, $s$, and secret $x \in \mathbb{Z}_q$:*

$$H_\infty(\text{output} \mid x) = -\log_2\left(\max_v \Pr[\text{output} = v \mid x]\right) \geq -\log_2\left(\frac{2}{q}\right) = \log_2(q) - 1$$

The min-entropy of the output distribution is at least $\log_2(q) - 1$ bits. For perfect masking (bijection), $H_\infty = \log_2(q)$. Barrett internal wires lose at most 1 bit compared to this ideal. Table 4 instantiates the bound for the two NIST standards.

| Standard | $q$ | $\log_2(q)$ | $H_\infty \geq$ | Leakage bound |
|---|---|---|---|---|
| ML-KEM (FIPS 203) | 3329 | 11.70 | 10.70 bits | ≤ 1 bit |
| ML-DSA (FIPS 204) | 8,380,417 | ≈ 22.99 | ≥ 21.99 bits | ≤ 1 bit |

**Table 4.** 1-Bit Barrier instantiated for the two NIST PQC standard moduli (FIPS 203 [5] and FIPS 204 [22]). Leakage is bounded above by 1 bit per Barrett internal wire under first-order arithmetic masking; the support-gap analysis (Table 3) shows actual leakage is often strictly less. We state Observations 5.1 and 5.2 as observations rather than formal theorems because Mathlib 4 does not currently define Shannon entropy or mutual information for finite types. The PFR project [23] provides these definitions but they have not been upstreamed to Mathlib. The algebraic content, count $\leq 2$, is the proved theorem (barrett_max_multiplicity_two). The information-theoretic interpretation follows by standard definitions. When Mathlib upstreams entropy machinery, the formal bound follows immediately from the proved algebraic bound.

### 6.3. Why This Matters for PQC Hardware

**For FIPS 140-3** [24] **certification.** A universal 1-bit bound enables evaluators to certify Barrett-based implementations without per-parameter-set re-verification. Any implementation satisfying the scope condition ($2^s \geq q$) inherits the bound. NIST IR 8547 [25] explicitly recommends formal methods for PQC implementation validation; the Lean 4 proof provides a machine-checked certificate that evaluators can verify independently.

**For design portability.** The bound holds for $q = 3329$ (ML-KEM), $q = 8{,}380{,}417$ (ML-DSA), and any future NIST PQC standard over any prime field. Post-standardization parameter updates do not invalidate the bound. A hardware IP block verified against the 1-Bit Barrier is portable across PQC standards without re-analysis.

**For tool trust.** The bound is machine-checked in Lean 4, not empirically observed, not assumed, not verified for specific instances only. The Lean kernel is the trusted computing base. This is a qualitatively different assurance level from empirical simulation or SAT-based checking.

## 7. Prime-Field PINI

### 7.1. The PF-PINI Definition

We define Prime-Field PINI (PF-PINI) as a direct preimage multiplicity bound for gadgets operating over $\mathbb{Z}_q$.

**Definition 6.1** (PF-PINI Gadget). *An PF-PINI gadget over* $\mathbb{Z}_q$ *is a structure consisting of: - A compute function* $G: \mathbb{Z}_q \times \mathbb{Z}_q \rightarrow \mathbb{Z}_q$ *mapping (secret, mask) to wire value - A max-multiplicity parameter* $k \in \mathbb{N}$ *- A proof that for all* $x, v \in \mathbb{Z}_q$*:* $|\{m: G(x, m) = v\}| \leq k$

In Lean 4:

```
structure PFPINIGadget (q : ℕ) [NeZero q] where
  /-- The gadget computation: (secret, mask) ↦ wire value -/
  compute : ZMod q → ZMod q → ZMod q
  /-- The PINI order parameter (max preimage size) -/
  maxMult : ℕ
  /-- Proof that every preimage has size ≤ maxMult -/
  bound : ∀ (x v : ZMod q),
    (univ.filter (fun m : ZMod q => compute x m = v)).card ≤ maxMult
```

*Reproduced verbatim from artifact; compiles against Lean 4.30.0-rc1 + Mathlib 322515540d7f.*

The parameter $k$ determines the security level: - $k = 1$: perfect masking (bijection), zero leakage. The butterfly satisfies this. - $k = 2$: 1-bit barrier, at most 1 bit leakage per wire. Barrett satisfies this. - $k > 2$: higher leakage. Not observed in PQC Barrett implementations.

**Relationship to standard notions.** PF-PINI($k$) is a preimage cardinality bound, not a simulation-based property. In the standard taxonomy: it does not directly correspond to $t$-NI, $t$-SNI, or PINI, all of which require a simulation argument (showing that any set of $t$ probes can be simulated from at most $t$ input shares). PF-PINI($k$) establishes a weaker but precisely quantified property: bounded max-probability per wire observation. For $k = 1$, PF-PINI(1) implies the algebraic core of 1-NI, per-context uniformity over the fresh mask, as shown in Paper 4 [4]. For $k = 2$, PF-PINI(2) establishes a bounded distinguishing advantage ($\leq 2/q$) per wire under the $t = 1$ probing model, but does not provide a simulation argument.

Boolean PINI [8] requires that any probe can be simulated using at most one share; PF-PINI($k$) does not claim this. The two notions are complementary: PF-PINI provides a universal machine-checked bound; PINI provides composable simulation guarantees.

### 7.2. Barrett is PF-PINI(2)

```
def barrettPFPINI {q : ℕ} [NeZero q] (s : ℕ) : PFPINIGadget q where
  compute := barrettInternalMap s
  maxMult := 2
  bound := barrett_max_multiplicity_two s
```

*Reproduced verbatim from artifact; compiles against Lean 4.30.0-rc1 + Mathlib 322515540d7f.*

Barrett reduction satisfies PF-PINI(2). The proof obligation `barrett_max_multiplicity_two s` is exactly Theorem 4.4, the 1-Bit Barrier. The instantiation is a direct application of the main theorem.

### 7.3. Butterfly is PF-PINI(1)

```
def identityPFPINI {q : ℕ} [NeZero q] : PFPINIGadget q where
  compute := fun x m => x - m
  maxMult := 1
  bound := by
    intro x v
    suffices h : (univ.filter (fun m : ZMod q => x - m = v)) = {x - v} by
      rw [h, card_singleton]
    ext m; simp only [mem_filter, mem_univ, true_and, mem_singleton]
    constructor
    · intro h; linear_combination -h
    · intro h; rw [h]; ring
```

*Reproduced verbatim from artifact; compiles against Lean 4.30.0-rc1 + Mathlib 322515540d7f.*

The identity masking gadget computes $x - m$ for a secret-dependent constant $x$ and fresh mask $m$. The preimage of any value $v$ is the singleton $\{x - v\}$, exactly one mask produces each output. This is the algebraic structure underlying Paper 4's butterfly_wire_count_eq_one [4]: each butterfly output wire reduces to the form $c - m$ for some constant $c$ depending on secrets and twiddle factors. The identityPFPINI construction proves PF-PINI(1) for this wire map directly; Paper 4 proves that every butterfly wire has this form. PF-PINI(1) means perfect uniformity: zero leakage beyond what the fresh mask provides.

The PF-PINI hierarchy: PF-PINI(1) ⊂ PF-PINI(2) ⊂ PF-PINI($k$) for $k \geq 2$. Every PF-PINI(1) gadget is also PF-PINI(2), but the converse is false. The butterfly is strictly better than Barrett under this framework.

### 7.4. The Composition Observation

**Observation 6.1** (Composition with fresh masking). *With fresh independent masks between pipeline stages:*

$$\text{max-multiplicity}(G_1 \circ G_2) = \max(k_1, k_2)$$

Without fresh masking:

$$\text{max-multiplicity}(G_1 \circ G_2) \leq k_1 \times k_2$$

This is an observation supported by the PF-PINI instantiations, not a proved theorem.

**Intuition.** With fresh masking, the intermediate value between stages is re-randomized. Stage 2 receives a uniformly masked input regardless of Stage 1's leakage. The stages decouple: each wire is independently bounded by its own PF-PINI parameter. The pipeline's overall bound per wire is the maximum over individual stages.

Without fresh masking, the $k_1$ output choices from Stage 1 each propagate $k_2$ output choices through Stage 2, giving $k_1 \times k_2$ in the worst case.

For the butterfly (PF-PINI(1)) → Barrett (PF-PINI(2)) pipeline with fresh masking:

$$\max(1,2) = 2$$

The 1-Bit Barrier propagates through the full NTT pipeline. Every wire, including those in Barrett reduction stages, leaks at most 1 bit of min-entropy.

The formal proof of this composition observation is future work. It requires a fresh-mask distribution argument extending Paper 4's ntt_pipeline_composition approach to the mixed PF-PINI(1)/PF-PINI(2) case.

## 8. Adams Bridge Connection

### *8.1 Two Distinct Failure Modes*

Adams Bridge [13] has two independent failure modes, now formally characterized across the five-paper program.

**Intra-stage failure (Papers 1 and 2).** Within each NTT stage, both shares interact through combinational logic without fresh intermediate masking. QANARY's structural dependency analysis [1] detects convergence points where $\text{share}_0$ and $\text{share}_1$ paths merge. Paper 1 found 8,487 convergence points across 30 modules and 165 INSECURE_CONSERVATIVE wires in the Barrett module alone. Paper 2's belief propagation attack [2] exploits these convergence points, achieving full key recovery under realistic noise assumptions. Concurrent CPA-attack work on the same Adams Bridge platform by Karabulut and Azarderakhsh [26] empirically confirmed the practical exploitability of the ML-DSA Barrett implementation.

**Inter-stage failure (this paper).** Adams Bridge enables masking (masking_en_ctrl = 1) only during rounds_count == 0 (ntt_ctrl.sv:264-272). NTT rounds 1–3 proceed without fresh per-stage masks. The PF-PINI composition bound (Observation 6.1) assumes fresh inter-stage masking. Adams Bridge violates this hypothesis. Therefore the $\max(k_1, k_2) = 2$ composition bound does not apply. Instead, the worst-case multiplicative bound $k_1 \times k_2$ applies to unmasked stage combinations. Over multiple rounds without fresh masking, the worst-case max-multiplicity is bounded by the product of per-stage parameters. For a pipeline with $r$ Barrett stages (each PF-PINI(2)) and butterfly stages (each PF-PINI(1)), this is at most $2^r$, which exceeds the 1-bit barrier.

### 8.2. The 165 Barrett Wires

Paper 1 [1] classified the ML-KEM Barrett module's 198 wires as 165 INSECURE_CONSERVATIVE and 33 SECURE. The trichotomy now provides a precise characterization of the 165 insecure wires.

Each insecure wire is an internal Barrett wire where QANARY's value-independence check could not confirm security, meaning the wire value depends on both shares. The trichotomy (Theorem 4.5) characterizes these wires: for each secret $x$ and output value $v$, the number of masks producing $v$ is 0, 1, or 2.

The support gap structure determines leakage as a function of the input. Every secret $x \neq q - 1$ has at least one unreachable output value. The number of unreachable values follows a triangular profile (Observation 4.1):

- For small $x$ (e.g., $x = 0$): 1 unreachable value. Nearly full support, worst-case 1-bit leakage.
- For mid-range $x$ (e.g., $x \approx q/2$): up to $q - r = 944$ unreachable values (ML-KEM). Actual leakage is noticeably less than 1 bit.
- At $x = q - 1 = 3328$ : the map is a bijection (count = 1 everywhere, zero unreachable). Zero leakage.

### 8.3 The Prescriptive Result

The 1-Bit Barrier tells hardware designers what Barrett costs and what to do about it.
Barrett reduction under first-order arithmetic masking with shift $s \geq \lceil \log_2 q \rceil$ leaks at most 1 bit of min-entropy per internal wire. This is the universal leakage floor: no masking scheme operating through Barrett's two-branch conditional subtraction can do better without changing the algorithm. A designer can achieve exactly this level by:

1. Using fresh per-stage masking between NTT butterfly and Barrett stages (Paper 4's Fresh Masking Design Principle [4]).
2. Ensuring $s \geq \lceil \log_2 q \rceil$ (trivially satisfied by all standard Barrett implementations).
3. Accepting 1 bit of worst-case internal leakage per Barrett wire.

Adams Bridge satisfies condition (2) but violates condition (1). That is the architectural root cause of its vulnerabilities across the five-paper program.

## 9. Proof Suite Summary

Table 5 enumerates every theorem and verified construction in the Lean 4 artifact accompanying this paper.

| # | Theorem | File | Key Technique | Sorry-Free | Closes |
|---|---|---|---|---|---|
| 1 | barrettInternalMap_eq_or | Basic | unfold + split | Yes | Two-branch structure |
| 2 | barrettInternalMap_mem_pair | Basic | eq_or case split + linear_combination | Yes | Preimage characterization |

| # | Theorem | File | Key Technique | Sorry-Free | Closes |
|---|---|---|---|---|---|
| 3 | barrettInternalMap_preimage_subset | BarrettUniversal | filter_subset | Yes | Formal preimage bound |
| 4 | barrett_max_multiplicity_two | BarrettUniversal | card_le_card + card_insert_le | Yes | **THE 1-BIT BARRIER** |
| 5 | barrett_multiplicity_trichotomy | BarrettUniversal | omega on Nat bounds | Yes | Complete structure |
| 6 | barrettInternalMap_candidate_A | BarrettUniversal | unfold + if_pos + ring | Yes | Branch A existence |
| 7 | barrettInternalMap_candidate_B | BarrettUniversal | unfold + if_neg + ring | Yes | Branch B existence |
| 8 | barrett_count_ge_one_of_A | BarrettUniversal | candidate membership | Yes | Lower bound (Branch A) |
| 9 | barrett_count_ge_one_of_B | BarrettUniversal | candidate membership | Yes | Lower bound (Branch B) |
| 10 | barrett_count_eq_two | BarrettUniversal | explicit witness + omega | Yes | Tightness (when both branches apply and $r \neq 0$) |
| 11 | barrettPFPINI | BarrettUniversal | barrett_max_multiplicity_two | Yes | Barrett PF-PINI(2) |
| 12 | identityPFPINI | BarrettUniversal | filter singleton + ring | Yes | Identity map PF-PINI(1) (butterfly wire form) |

**Table 5.** Complete proof suite: 10 theorems + 2 verified PF-PINI gadget constructions = 12 results, all kernel-verified with zero sorry. See the *Code and Data Availability* section for repository, toolchain, and reproduction instructions.

## 10. Limitations and Future Work

We state all limitations explicitly.

### 10.1. (i) Support gap formula not proved

The zero-multiplicity region size, $\min(x.\text{val}+1,\ q-r,\ q-1-x.\text{val})$ where $r = 2^s \bmod q$, is computationally verified across 14 primes (including all PQC standard moduli: $q = 3329$, $q = 8{,}380{,}417$, $q = 7681$, $q = 4591$, $q = 12289$) but is not a proved Lean theorem. The `ZMod.val` arithmetic required is technically tractable but not yet formalized. This is the highest-priority future proof target: it would convert the support gap observation into a machine-checked characterization and enable a tighter, secret-dependent min-entropy bound.

### 10.2. (ii) PF-PINI composition not proved

The $\max(k_1, k_2)$ composition bound with fresh masking (Observation 6.1) is an observation, not a theorem. The formal proof requires tracking how the fresh intermediate mask distributes the output of Stage 1 uniformly to Stage 2. This extends Paper 4's `ntt_pipeline_composition` argument to the mixed PF-PINI(1)/PF-PINI(2) case and is research-level work: the fresh-mask distribution argument interacts with the PF-PINI preimage bound in a way that requires new infrastructure.

### 10.3. (iii) Information-theoretic formalization

The min-entropy bound $H_\infty \geq \log_2(q) - 1$ (Observation 5.2) is an informal consequence of the proved algebraic bound. Mathlib 4 does not currently define Shannon entropy or mutual information for finite types. Paper 3's `MutualInfoZero` [3] provides an algebraic proxy for the zero-MI case. When Mathlib upstreams the PFR project's [23] entropy machinery, the formal bound follows immediately from the proved algebraic bound `barrett_max_multiplicity_two`. Until then, the gap between the proved algebraic bound and the information-theoretic statement is bridged by standard definitions.

### 10.4. (iv) Higher-order masking

All results are first-order: one probe, two shares. At second order ($d = 2$ probes, three shares), the two-branch structure means an adversary probing two carefully chosen wires could potentially exploit the Branch A / Branch B overlap. Each individual probe sees at most 2 preimage values, but two simultaneous probes may interact. The higher-order Barrett security analysis requires a different preimage characterization, joint preimage sets across multiple wires, and is future work with significant technical challenges.

### 10.5. (v) Non-Barrett modular reduction

The results apply specifically to the Barrett algorithm with shift $s$. Montgomery reduction uses a different decomposition (multiplication by $q^{-1} \bmod 2^s$ and right-shift), producing a different internal wire structure. Schoolbook reduction ($x - \lfloor x/q \rfloor \cdot q$) has yet another structure. Each scheme requires separate analysis. The PF-PINI framework itself is general, but the max-multiplicity bound for each scheme must be proved individually.

### 10.6. (vi) Nat equivalence bridge

The `barrettInternalMapNat` definition (hardware-faithful, using Nat modular arithmetic) is computationally equivalent to `barrettInternalMap` (the algebraic two-branch definition used for proofs). The equivalence is defined in the artifact but not proved. This is a technical gap for connecting the Lean theorem to the actual Adams Bridge RTL: the theorem applies to the algebraic definition, and the hardware implements the Nat arithmetic. The computational verification (all $3329^2$ input pairs match) provides strong evidence but is not a kernel-verified proof.

### 10.7. Future work summary

Ordered by priority: 1. Support gap formula proof (would tighten the bound) 2. PF-PINI composition theorem (would complete the pipeline argument) 3. Entropy formalization (when Mathlib provides the infrastructure) 4. Higher-order Barrett analysis (different mathematical regime) 5. Montgomery/schoolbook reduction characterization (extends the framework) 6. Nat equivalence bridge (connects theorem to hardware)

## 11. Conclusion

We have proved three machine-checked results for arithmetic masking in Barrett reduction, formalized in Lean 4 with Mathlib, with zero sorry.

**The trichotomy.** The Barrett internal wire map has preimage cardinality in $\{0,1,2\}$, universal over all $q > 0$ and all shift parameters $s$. The bound is tight: count = 2 is achievable. The count-zero cases reduce the adversary's information, making the 1-Bit Barrier a conservative bound.

**The 1-Bit Barrier.** Max-multiplicity 2 implies $H_\infty \geq \log_2(q) - 1$ per internal wire. Every Barrett implementation over any modulus, with any shift parameter, leaks at most 1 bit of min-entropy per internal wire under first-order probing. The kernel-checked content is the cardinality bound card $\leq 2$; the 1-bit min-entropy interpretation follows by standard definitions (Limitation (iii)). This is the first universal machine-checked cardinality bound for Barrett reduction, from which the corresponding leakage bound follows.

**The PF-PINI framework.** Barrett satisfies PF-PINI(2); the butterfly satisfies PF-PINI(1). We observe (not yet proved) that with fresh inter-stage masking, the composed pipeline inherits the maximum per-stage parameter, the 1-Bit Barrier propagates.

The proof is clean because of the two-branch algebraic structure of modular reduction. The three-step argument, preimage is a subset of two candidates, each candidate contributes at most once, so the preimage has at most two elements, is not complex. It is inevitable once you see the structure. This is the pattern across the program: the right algebraic abstraction makes the result obvious. The contribution is not the mathematical difficulty. It is that nobody characterized Barrett's preimage structure, nobody proved the universal leakage bound, and nobody provided hardware designers with a machine-checked, universally-quantified result they can cite in certification documentation.

Two independent failure modes in Adams Bridge are now formally characterized. Intra-stage failures (Papers 1 [1] and 2 [2]): both shares interact through combinational logic without intermediate masking, creating convergence points exploitable by belief propagation and structural analysis. Inter-stage failure (this paper): the absence of fresh per-stage masking violates the PF-PINI composition hypothesis, so the $\max(k_1, k_2)$ bound does not apply.

Together, these give the complete architectural picture of Adams Bridge's security properties under the first-order probing model.
The five-paper program: Paper 1 [1] built the detection tool. Paper 2 [2] quantified the attack. Paper 3 [3] proved universal algebraic foundations. Paper 4 [4] proved butterfly composition. Paper 5 proved the 1-Bit Barrier for Barrett. The arc is complete. The verified toolchain, from mathematical design principle through universal leakage bounds to gate-level certification evidence, now exists for arithmetic-masked NTT hardware.
Any Barrett reduction implementation over any modulus with shift parameter $s$ has max-multiplicity at most 2 per internal wire under first-order probing, machine-checked and universal. The resulting 1-bit min-entropy bound follows from standard definitions and is ready for FIPS 140-3 certification arguments.

## 12. Code and Data Availability

The Lean 4 artifact accompanying this paper is publicly available under the MIT license:

- **Repository:** `https://github.com/rayiskander2406/qanary-one-bit-barrier-arXiv-2604.24670` (arXiv ID will be substituted post-submission per series convention)
- **Toolchain:** Lean 4 v4.30.0-rc1 (managed via `elan`)
- **Pinned dependency:** Mathlib at commit `322515540d7fd29ef8992b82c89044f86f02ac10`
- **License:** MIT (artifact); CC-BY-4.0 (manuscript)

### 12.1. Reproduction

```
git clone https://github.com/rayiskander2406/qanary-one-bit-barrier-arXiv-2604.24670
cd qanary-one-bit-barrier-arXiv-2604.24670
lake build            # ~30 min on first run; downloads + compiles Mathlib
python3 reproduce.py --check
```

Expected output: 12 proved results, zero `sorry`, zero `admit`, zero added `axiom`, zero `native_decide` calls, zero errors. The `reproduce.py --check` script verifies the theorem-name index in `QanaryPaper5/Basic.lean` and `QanaryPaper5/BarrettUniversal.lean` matches the proof suite enumerated in Section 8.

### 12.2. Artifact contents

Table 6 maps each top-level path in the artifact to its purpose.

| Path | Purpose |
|---|---|
| QanaryPaper5/Basic.lean | Barrett internal wire map definitions (barrettInternalMap, barrettInternalMapNat); barrettInternalMap_eq_or and barrettInternalMap_mem_pair |
| QanaryPaper5/BarrettUniversal.lean | Trichotomy theorem, 1-Bit Barrier, |

| Path | Purpose |
|---|---|
| | tightness witness, PF-PINI(2)/PF-PINI(1) instantiations (10 theorems + 2 verified constructions) |
| lakefile.lean, lake-manifest.json | Build configuration with pinned dependency revisions |
| lean-toolchain | Pins leanprover/lean4:v4.30.0-rc1 |
| LICENSE, CITATION.cff | MIT license and machine-readable citation metadata |
| reproduce.py | Single-command verification entry point (--check mode skips re-build) |
| README.md, DESIGN.md | Top-level documentation, design rationale, theorem index |

**Table 6.** Artifact path inventory. The two .lean source files contain all definitions and proofs enumerated in Table 5.

The artifact is self-contained: it depends only on Mathlib and does not import from the Paper 3 [3] or Paper 4 [4] artifacts. Cross-references to those works in the manuscript are conceptual.

### 12.3. Archival DOI

The Lean 4 artifact is archived on Zenodo for long-term preservation:

- **Concept DOI** (always resolves to the latest version, cite this for "the artifact"): 10.5281/zenodo.19842166
- **v1.0.1 version DOI** (fixed to commit `a0207b1`, for bit-for-bit reproducibility pinning): 10.5281/zenodo.19842167

### 12.4. Independent reproducibility

The artifact has no external data dependencies, no random seeds, and no networked services beyond the initial lake package fetch. Verification is purely a function of the Lean 4 kernel and the pinned Mathlib commit; identical inputs yield identical outputs across machines. The trusted computing base is the Lean 4 kernel; this paper's artifact contains no native_decide invocations.